\documentclass[12pt]{article}
\usepackage{bm,amsmath,amssymb,graphicx}

\begin{document}
\pagenumbering{arabic}
\begin{titlepage}

\title{Singularities in massive conformal gravity}

\author{F. F. Faria$\,^{*}$ \\
Centro de Ci\^encias da Natureza, \\
Universidade Estadual do Piau\'i, \\ 
64002-150 Teresina, PI, Brazil}

\date{}
\maketitle

\begin{abstract}
We study the quantum effects of big bang and black 
hole singularities on massive conformal gravity. We do 
this by analyzing the behavior of the on-shell effective 
action of the theory at these singularities. The result 
is that such singularities are harmless in massive conformal 
gravity because the on-shell effective action of the theory 
does not diverge at them.
\end{abstract}

\thispagestyle{empty}
\vfill
\bigskip
\noindent * felfrafar@hotmail.com \par
\end{titlepage}
\newpage


\section{Introduction}
\label{sec1}


In relativistic theories of gravity, such as general relativity (GR), the 
term ``singularity" can be used to describe a point in spacetime 
where the curvature of spacetime becomes infinite, which is associated with 
such spacetime containing incomplete geodesics \cite{Pen1,Haw}. The 
singularities usually arise in cosmological and black hole 
spacetime solutions of most relativistic theories of gravity. Therefore, 
studying them could lead to advances in our understanding of several 
unexplained physical phenomena, such as the origin of the universe or the 
black hole information paradox, to name just two. Unfortunately, both the 
infinity of spacetime curvature and the geodesic incompleteness make it 
impossible to predict any classical physical behavior in the regions near the 
singularities. However, in these regions the gravitational field gets so strong 
that quantum effects become dominant, which means that the singularities have 
effects only on the quantum observables of a relativistic theory of gravity, such 
as the $S$-matrix elements, and not on the classical ones.

A fundamental quantity for the calculation of the quantum observables of a 
field theory is the on-shell effective action, which can be found through 
the perturbation expansion
\begin{equation}
\Gamma[\Phi_\mathrm{cl}] \approx S[\Phi_\mathrm{cl}] 
+ \frac{i}{2}\mathrm{Tr}\left[\log{\frac{\delta^2 S[\Phi]}
{\delta \Phi(x)\delta \Phi(y)}}\bigg|_{\Phi=\Phi_\mathrm{cl}}\right] 
+ \cdots,
\label{1}
\end{equation}
where $\Phi_\mathrm{cl}$ represents the dynamical fields that are solutions 
of the classical field equations of the theory and $S[\Phi_\mathrm{cl}]$ is 
the on-shell classical action of the theory. A consistent on-shell effective 
action requires that either all of its expansion terms be finite or that 
its classical term be finite and its perturbed terms be at least renormalizable. 
In this way, there is no consistency problem with the classical solutions of a 
renormalizable field theory being singular as long as the on-shell classical 
action of the theory does not diverge at the singularity. This idea is in 
line with the finite action principle first proposed in Ref. \cite{Bar1} 
and recently revisited in Ref. \cite{Bar2}.

Unfortunately, GR is not perturbatively renormalizable \cite{Gor,Ven} and 
thus we cannot analyze the quantum effects of the curvature singularities on 
it\footnote{Here we are not judging the validity of GR at low energies 
as an effective theory. What we are saying is that even though the theory has 
singular spacetimes whose on-shell classical actions are finite, such as the 
Schwarzschild spacetime, its on-shell effective action is divergent at two-loop 
level and thus the theory cannot solve the singularity problem, which is a 
quantum problem and not a classical one.}. Instead, here we will study such 
effects in a power-counting renormalizable \cite{Far1,Far2} 
theory of gravity called massive conformal gravity (MCG) \cite{Far3}. Besides 
being invariant under coordinate transformations, this theory is also invariant 
under the (local) conformal transformations
\begin{equation}
\tilde{\Phi} = \Omega(x)^{-\Delta_{\Phi}}\Phi,
\label{2}
\end{equation}
where the conformal factor $\Omega(x)$  is an arbitrary function of the 
spacetime coordinates, and $\Delta_{\Phi}$ is the scaling dimension of 
the dynamical field $\Phi$, whose values are $-2$ for the metric field, 
$0$ for gauge bosons, $1$ for scalar fields and $3/2$ for fermions. 
The extra conformal symmetry of MCG allows us to move the ghost pole 
that appears in the massive spin-$2$ propagator of the theory by varying 
the conformal gauge fixing parameter such that the ghost pole does not 
contribute to the gauge-invariant absorptive part of the $S$-matrix, 
which leads to the unitarity of the theory \cite{Far4}. This gives to 
MCG an advantage over other renormalizable theories of gravity that are 
non-unitary such as higher-derivative gravity \cite{Ste,John}.

The unitarity of MCG is a crucial factor in choosing it for the study of 
the quantum effects of the singularities, since a non-unitary $S$-matrix is 
unphysical. Furthermore, classically\footnote{The ghost can 
also produce instabilities in the classical solutions of the theory. However, 
the solutions studied so far have proven to be stable.}, it has been shown 
so far that MCG is free of the van Dam-Veltman-Zakharov (vDVZ) discontinuity 
\cite{Far5}, it can reproduce the orbit of binaries by the emission of 
gravitational waves \cite{Far6}, it fits well with the Type Ia supernovae 
(SNIa) data without the cosmological constant problem \cite{Far7}, it is 
consistent with solar system observations \cite{Far8} and it predicts the 
observed primordial abundances of light elements \cite{Far9}.

This paper is organized as follows. In the next section, 
we derive the MCG on-shell classical total action. In Sec. \ref{sec3}, we 
present a detailed discussion on the renormalizability of MCG. In Sec. 
\ref{sec4}, we analyze the quantum effects of the big bang singularity on 
the theory. In Sec. \ref{sec5}, we repeat the same analysis made in the 
previous section but now for the case of a black hole singularity. Finally, 
conclusions are given in Sec. \ref{sec6}.     


\section{On-shell classical action}
\label{sec2}


The total conformally invariant action of MCG is given 
by\footnote{Here we use units in which $c = \hbar = 1$.}
\begin{eqnarray}
S &=& S_{\textrm{MCG}} + S_{m} \nonumber \\
&=&\int{d^{4}x} \, \sqrt{-g}\bigg[ \varphi^{2}R 
+ 6 \partial^{\mu}\varphi\partial_{\mu}\varphi 
- \frac{1}{2\alpha^2} C^2 \bigg] 
+ \int{d^{4}x\mathcal{L}_{m}},
\label{3}
\end{eqnarray}
where $\varphi$ is a gravitational scalar field called 
dilaton, $\alpha$ is a dimensionless coupling constant, 
\begin{equation}
C^2 = C^{\alpha\beta\mu\nu}C_{\alpha\beta\mu\nu} = R^{\alpha\beta\mu\nu}
R_{\alpha\beta\mu\nu} - 2R^{\mu\nu}R_{\mu\nu} + \frac{1}{3}R^{2}
\label{4}
\end{equation}
is the Weyl curvature invariant,
$R^{\alpha}\,\!\!_{\mu\beta\nu} 
= \partial_{\beta}\Gamma^{\alpha}_{\mu\nu} + \cdots$ is the Riemann tensor, 
$R_{\mu\nu} = R^{\alpha}\,\!\!_{\mu\alpha\nu}$ is the Ricci tensor, 
$R = g^{\mu\nu}R_{\mu\nu}$ is the scalar curvature, and 
$\mathcal{L}_{m} = \mathcal{L}_{m}[g_{\mu\nu},\Psi]$\footnote{We must 
consider that the dilaton field does not couple with matter to provide 
renormalizability, as we will show in Sec. \ref{sec3}.} is the Lagrangian 
density of the matter field $\Psi$.   

Varying the action (\ref{3}) with respect to $g^{\mu\nu}$ and $\varphi$, 
we obtain the MCG field equations
\begin{equation}
\varphi^{2}G_{\mu\nu} +  6 \partial_{\mu}\varphi\partial_{\nu}\varphi 
- 3g_{\mu\nu}\partial^{\rho}\varphi\partial_{\rho}\varphi + g_{\mu\nu} 
\Box \varphi^{2} 
- \nabla_{\mu}\nabla_{\nu} \varphi^{2}  - \alpha^{-2} B_{\mu\nu} 
= \frac{1}{2}T_{\mu\nu},
\label{5}
\end{equation}
\begin{equation}
\left(\Box - \frac{1}{6}R \right) \varphi = 0,
\label{6}
\end{equation}
where
\begin{equation}
B_{\mu\nu} = \Box R_{\mu\nu} 
- \frac{1}{3}\nabla_{\mu}\nabla_{\nu}R  -\frac{1}{6}g_{\mu\nu}\Box 
R + 2R^{\rho\sigma}R_{\mu\rho\nu\sigma} 
-\frac{1}{2}g_{\mu\nu}R^{\rho\sigma}R_{\rho\sigma}  
- \frac{2}{3}RR_{\mu\nu}  + \frac{1}{6}g_{\mu\nu}R^2
\label{7}
\end{equation}
is the Bach tensor,
\begin{equation}
G_{\mu\nu} = R_{\mu\nu} - \frac{1}{2}g_{\mu\nu}R
\label{8}
\end{equation}
is the Einstein tensor,
\begin{equation}
\Box \varphi = 
\frac{1}{\sqrt{-g}}\partial^{\rho}\left( \sqrt{-g} \partial_{\rho}
\varphi \right)
\label{9}
\end{equation} 
is the generally covariant d'Alembertian for a scalar field,
\begin{equation}
T_{\mu\nu} = - \frac{2}{\sqrt{-g}} \frac{\delta \mathcal{L}_{m}}
{\delta g^{\mu\nu}}
\label{10}
\end{equation}
is the matter energy-momentum tensor, and $\Box 
= \nabla^{\mu}\nabla_{\mu}$ .

A particular feature of MCG is that both its line element
\begin{equation}
ds^2 = \left( \varphi/\varphi_{0}\right)^2 g_{\mu\nu} dx^{\mu}dx^{\nu}
\label{11}
\end{equation}
and its geodesic equation
\begin{equation}
\frac{d^{2}x^{\lambda}}{d\tau^2} + \Gamma^{\lambda}\,\!\!_{\mu\nu}
\frac{dx^{\mu}}{d\tau}\frac{dx^{\nu}}{d\tau} +\frac{1}{\varphi}
\frac{\partial\varphi}{\partial x^{\rho}} \left( g^{\lambda\rho} + 
\frac{dx^{\lambda}}{d\tau}\frac{dx^{\rho}}{d\tau}\right) = 0
\label{12}
\end{equation}
are also conformally invariant, where $\varphi_0$ is a constant dilaton 
field and
\begin{equation}
\Gamma^{\lambda}\,\!\!_{\mu\nu} = \frac{1}{2}g^{\lambda\rho}\left( 
\partial_{\mu}g_{\nu\rho} + \partial_{\nu}g_{\mu\rho} 
- \partial_{\rho}g_{\mu\nu} \right)
\label{13}
\end{equation}
is the Levi-Civita connection.

Noting that the dilaton field $\varphi$ acquires the spontaneously 
broken vacuum expectation value $\varphi_{0}$ below the Planck scale 
\cite{Matsuo}, we find that the MCG action becomes
\begin{equation}
S_{\textrm{MCG}}[\varphi_0, g_{\mu\nu}] = \int{d^{4}x} \, 
\sqrt{-g}\bigg[ \varphi_{0}^{2}R  
- \frac{1}{2\alpha^2} C^2 \bigg]
\label{14}
\end{equation}
at the classical level, with the metric $g_{\mu\nu}$ being now the only classical 
dynamical gravitational field of the theory. The variation of (\ref{14}) plus 
the matter action $S_m$ with respect to $g^{\mu\nu}$ gives the classical MCG field 
equations\footnote{Actually, the field equation (\ref{16}) is obtained by tanking 
the trace of (\ref{15}) and using the fact that just like the Bach tensor 
$B_{\mu\nu}$, the energy-momentum tensor  $T_{\mu\nu}$ is also traceless in 
conformally invariant theories such as MCG. Alternatively, we can find the field 
equations (\ref{15}) and (\ref{16}) in an easier way by substituting 
$\varphi = \varphi_0$ into (\ref{5}) and (\ref{6}). Making this same substitution 
into (\ref{11}), we can also find that the classical MCG line element is given by 
$ds^2 = g_{\mu\nu} dx^{\mu}dx^{\nu}$.}
\begin{equation}
\varphi_{0}^{2}G_{\mu\nu} - \alpha^{-2} B_{\mu\nu} 
= \frac{1}{2}T_{\mu\nu},
\label{15}
\end{equation}
\begin{equation}
R = 0.
\label{16}
\end{equation}
Finally, using (\ref{16}) in (\ref{14}), and considering the relation (\ref{4}), 
we obtain the on-shell classical MCG action
\begin{equation}
S^{\textrm{on-shell}}_{\textrm{MCG}} = 
- \frac{1}{2\alpha^2} \int{d^{4}x} \, \sqrt{-g}\bigg[ 
R^{\alpha\beta\mu\nu}R_{\alpha\beta\mu\nu} - 2R^{\mu\nu}R_{\mu\nu} \bigg],
\label{17}
\end{equation}
which need to be finite at the singularities for the theory to be consistent 
at the quantum level, as shown in the previous section.  

In fact, the complete quantum consistency of MCG requires that its 
on-shell classical total action $S^{\textrm{on-shell}} = 
S^{\textrm{on-shell}}_{\textrm{MCG}} + S^{\textrm{on-shell}}_{m}$ does not 
diverge at the singularities, where $S^{\textrm{on-shell}}_{m}$ is the 
on-shell classical matter action. This will happen if both 
$S^{\textrm{on-shell}}_{\textrm{MCG}}$ and $S^{\textrm{on-shell}}_{m}$ are 
finite or if the divergences in them cancel each other out. One way or 
another, we need to find the on-shell classical matter action to 
analyze all the quantum effects that the singularities have on the theory. 

Without loss of generality for our purposes, we can
consider the conformally invariant matter Lagrangian density\footnote{We do 
not include gauge bosons in (\ref{18}) because 
they do not contribute to the singular solutions studied here.} \cite{Man1}
\begin{equation}
\mathcal{L}_{m} = - \sqrt{-g}\bigg[S^{2}R + 6\partial^{\mu}S\partial_{\mu}S 
+ \lambda S^{4} + \frac{i}{2}\left(\, \overline{\psi}
\gamma^{\mu}D_{\mu}\psi - D_{\mu}\overline{\psi}\gamma^{\mu}\psi \right) 
- \mu S\overline{\psi}\psi\bigg],
\label{18}
\end{equation}
where $S$ is a scalar Higgs field, $\lambda$ and $\mu$ are dimensionless 
coupling constants, $\overline{\psi} = \psi^{\dagger}
\gamma^{0}$ is the adjoint fermion field, $D_{\mu} = \partial_{\mu} 
+ [\gamma^{\nu},\partial_{\mu}\gamma_{\nu}]/8 - [\gamma^{\nu},\gamma_{\lambda}]
\Gamma^{\lambda}\,\!\!_{\mu\nu}/8$, and $\gamma^{\mu}$ 
are the general relativistic Dirac matrices, which satisfy the anti-commutation 
relation $\{\gamma^{\mu},\gamma^{\nu}\} = 2g^{\mu\nu}$.

Considering that, at scales below the electroweak scale, the Higgs field 
acquires a spontaneously broken constant vacuum expectation value $S_{0}$, we 
find that (\ref{18}) becomes
\begin{equation}
\mathcal{L}_{m}[S_0, g_{\mu\nu}, \psi] = - \sqrt{-g}\bigg[S_0^{2}R  
+ \lambda S_{0}^{4} + \frac{i}{2}\left(\, \overline{\psi}\gamma^{\mu}D_{\mu}\psi 
- D_{\mu}\overline{\psi}\gamma^{\mu}\psi \right) 
- m_{\psi}\overline{\psi}\psi\bigg]
\label{19}
\end{equation}
at the classical level, where $m_{\psi} = \mu S_{0}$ is the fermion mass. By 
varying (\ref{19}) with respect to $\overline{\psi}$ and $\psi$, we obtain the 
field equations
\begin{equation}
i\gamma^{\mu}D_{\mu}\psi - m_{\psi} \psi = 0,
\label{20}
\end{equation}
\begin{equation}
iD_{\mu}\overline{\psi}\gamma^{\mu} + m_{\psi} \overline{\psi} = 0.
\label{21}
\end{equation}
In addition, the substitution of (\ref{19}) into (\ref{10}) gives
\begin{eqnarray}
T_{\mu\nu} &=& 2S_{0}^{2}G_{\mu\nu}
- g_{\mu\nu}\left[\lambda S_0^4 + \frac{i}{2}\left(\, \overline{\psi}
\gamma^{\rho}D_{\rho}\psi - D_{\rho}\overline{\psi}\gamma^{\rho}\psi \right) 
- m_{\psi}\overline{\psi}\psi\right] \nonumber \\ &&
+ \, \frac{i}{4}\big(\, \overline{\psi}
\gamma_{\mu}D_{\nu}\psi - D_{\nu}\overline{\psi}\gamma_{\mu}\psi 
+ \overline{\psi}\gamma_{\nu}D_{\mu}\psi - D_{\mu}\overline{\psi}\gamma_{\nu}
\psi \big).
\label{22}
\end{eqnarray}

Finally, using the trace of (\ref{22}) and the field equations (\ref{15}), 
(\ref{16}), (\ref{20}) and (\ref{21}) in (\ref{19}), we find the on-shell 
classical matter action
\begin{equation}
S^{\textrm{on-shell}}_{m} = 
- \frac{1}{4} \int{d^{4}x} \, \sqrt{-g}\bigg[ \frac{i}{2}\big(\, \overline{\psi}
\gamma^{\mu}D_{\mu}\psi - D_{\mu}\overline{\psi}\gamma^{\mu}\psi \big) \bigg],
\label{23}
\end{equation}
which can be written in the simplest form
\begin{equation}
S^{\textrm{on-shell}}_{m} = 
- \frac{1}{4} \int{d^{4}x} \, \sqrt{-g} \, T^{f},
\label{24}
\end{equation}
where $T^{f} = g^{\mu\nu}T^{f}_{\mu\nu}$ is the trace of the standard kinematic
fermion energy-momentum tensor\footnote{It is worth noting that the kinematic 
energy-momentum tensor (\ref{25}) contains only the energy of the fermion 
field while the dynamical energy-momentum tensor (\ref{22}) contains, in 
addition to the energy of the fermion field, the energy of the Higgs field 
that gives the mass of the fermion field and energy which couples to gravity. 
The dynamical mass generation of the fermion field makes the trace of 
(\ref{22}) zero. On the other hand, as (\ref{25}) is kinematic, its trace is
non-zero.}
\begin{equation}
T^{f}_{\mu\nu} = \frac{i}{4}\big(\, \overline{\psi}
\gamma_{\mu}D_{\nu}\psi - D_{\nu}\overline{\psi}\gamma_{\mu}\psi 
+ \overline{\psi}\gamma_{\nu}D_{\mu}\psi - D_{\mu}\overline{\psi}\gamma_{\nu}
\psi \big).
\label{25}
\end{equation}

In conclusion, we need to analyze the behavior of both 
(\ref{17}) and (\ref{24}) at the singularities to check the quantum consistency 
of MCG, what will be done in Sec. \ref{sec4} and Sec. \ref{sec5}. But before 
that, let us discuss about the renormalizability of the theory in the next 
section.


\section{Renormalizability}
\label{sec3}


As seen in Sec. {\ref{sec1}}, all of the analysis done in the previous section 
is meaningless if the perturbed terms of the MCG on-shell effective total action 
are not at least renormalizable. To check this, we note that the 
coordinate and conformal symmetries of the theory restrict the 
one-loop divergent term of the MCG effective action to be of the 
form\footnote{Since the theory is power-counting renormalizable, divergent 
terms with derivatives of order higher than four will not arise at any loop 
level.}
\begin{eqnarray}
\Gamma^{(1)}_{\textrm{MCG}} &=&
-\int{d^{4}x} \, \sqrt{-g} 
\Bigg[ \beta_1 C^2 + \beta_2 E + \beta_3
\left(\varphi^2R-6\varphi\Box\varphi\right)^2  \nonumber \\ && 
+ \, \beta_4\left(\varphi^2R-6\varphi\Box\varphi\right)
+ \beta_5 \varphi^4 \Bigg],
\label{26}
\end{eqnarray}
where $E = R^{\alpha\beta\mu\nu}R_{\alpha\beta\mu\nu}- 4R^{\mu\nu}R_{\mu\nu} 
+ R^2 $ is the Euler density, and $\beta_1$, $\beta_2$, $\beta_3$, $\beta_4$ 
and $\beta_5$ are some divergent coefficients.  

Using the field equation (\ref{6}) in (\ref{26}), we obtain the 
on-shell one-loop divergent term of the MCG effective action
\begin{equation}
\Gamma^{(1)\textrm{on-shell}}_{\textrm{MCG}} =
- \int{d^{4}x} \, \sqrt{-g} 
\left[ \beta_1 C^2 + \beta_2 E + \beta_5 \varphi^4  \right].
\label{27}
\end{equation}
Since the first term of (\ref{27}) is of the same type presented 
in the original MCG total action (\ref{3}), it is renormalizable. To 
renormalize the second term of (\ref{27}) we must add to (\ref{3}) 
the Gauss-Bonnet term  $\eta\int{d^{4}x} \, \sqrt{-g} E$. This can be done 
without problem because the Gauss-Bonnet term is a topological invariant 
that does not contribute to the field equations of the theory. The 
renormalization of the third term of (\ref{27}) is a little trickier. To do 
this, we must add to (\ref{3}) a quartic self-interacting term of the 
dilaton field $\lambda\int{\sqrt{-g} \varphi^4}$, which makes the flat 
metric no longer a solution of the field equations and, consequently,
invalidates the $S$-matrix formulation. However, we can solve this problem by 
considering the renormalized value of $\lambda$ equal zero so that the 
self-interacting term is present in the renormalized action only to cancel 
out divergent terms like the last one of (\ref{27}). In this way, we can 
consider that MCG is one-loop renormalizable. Additionally, it can be shown 
that the contribution of the conformally invariant matter fields to the 
on-shell one-loop divergent term of the MCG effective total action are of the 
same types presented in (\ref{3}) \cite{Buch}, which means that the total 
theory is one-loop renormalizable. 

Inserting the on-shell MCG effective action up to one-loop into 
the trace of the energy-momentum tensor
\begin{equation}
T = g^{\mu\nu}T_{\mu\nu} = -\frac{2}{\sqrt{-g}}g^{\mu\nu}\frac{\delta 
\Gamma}{\delta g^{\mu\nu}},
\label{28}
\end{equation}
we find that MCG has the conformal (trace) anomaly
\begin{equation}
T = \frac{1}{(4\pi)^2}\left[\beta_1 C^2 + \beta_2 E + \beta_5 \varphi^4 \right] 
\neq 0,
\label{29}
\end{equation}
which breaks the conformal symmetry of the theory at the one-loop level. 
A possible consequence of this symmetry breaking is the emergence of 
non-renormalizable $\int{\sqrt{-g}R^2}$, Einstein-Hilbert $\int{\sqrt{-g}R}$ 
and cosmological constant $\int{\sqrt{-g}\Lambda}$ divergent terms beyond 
the one-loop level. However, by performing the classical conformal 
transformations\footnote{This procedure can be regarded as a particular 
case of the conformal regularization method \cite{Frad} with the conformal 
factor given by $\Omega = \varphi/\varphi_{0}$.} 
\begin{equation}
\tilde{g}_{\mu\nu} = \left(\varphi/\varphi_{0}\right)^2 g_{\mu\nu}, 
\ \ \ \ \ \ \ \tilde{\varphi} = \varphi_{0},
\label{30}
\end{equation}
we can turn these non-renormalizable terms into the same types as the 
last three terms in (\ref{26}), which are either eliminated by the use 
of the field equation (\ref{6})\footnote{It is for the field equation 
(\ref{6}) to remain valid even in the presence of matter that we assumed that 
the dilaton field does not couple with matter in Sec. \ref{sec2}.} or 
renormalized, as we saw in the previous paragraph. This same procedure can be 
used to eliminate the non-renormalizable divergences that may arise beyond the 
one-loop level due to the conformal anomaly produced by the contribution of 
the conformally invariant matter fields. Thus, 
despite MCG having a conformal anomaly, all the 
perturbed divergent terms of its on-shell effective total action are 
renormalizable.


\section{Big bang singularity}
\label{sec4}


As stated earlier, MCG has only two known singular solutions with 
physical relevance. One of them is the MCG universe solution, whose 
geometry is described by the line element\footnote{It is worth noting that 
this solution is valid in all epochs of the MCG universe. This is because the 
dynamical mass generation of the fermion field causes the MCG universe to be 
filled with a dynamical perfect fluid whose energy
density decreases as $\rho \propto a^{-4}$ regardless of whether the particles 
that make up the fluid are relativistic or non-relativistic. Additionally, it 
is also worth noting that only the open MCG universe is consistent with 
cosmological observations so that (\ref{31}) is the only physical cosmological 
solution to the MCG field equations. Therefore, there is no point in analyzing 
the singularities of other cosmological solutions here.} 
\cite{Far7}
\begin{equation}
ds^{2} = - dt^{2} + a(t)^2\left( \frac{dr^2}{1+r^2}+ r^{2}d\theta^{2} 
+ r^{2}\sin^{2}\theta d\phi^{2} \right),
\label{31}
\end{equation}
where $a(t) = \sqrt{bt+t^2}$, with $b$ being a positive constant. Using this 
metric, we find the curvature invariants
\begin{eqnarray}
R^{\alpha\beta\mu\nu}R_{\alpha\beta\mu\nu} &=& \frac{24bt^2}{(t^2+bt)^4}, 
\label{32} \\
R^{\mu\nu}R_{\mu\nu} &=& \frac{12bt^2}{(t^2+bt)^4},
\label{33}
\end{eqnarray}
such that the spacetime of the MCG universe has a big bang singularity 
at $t=0$. 

To see the quantum effects of such singularity on the theory, we 
first substitute (\ref{32}) and (\ref{33}) into the on-shell classical MCG 
action (\ref{17}), which gives
\begin{equation}
S^{\textrm{on-shell}}_{\textrm{MCG}} = 0.
\label{34}
\end{equation}
Then, substituting the trace of the standard kinematic perfect fluid 
energy-momentum tensor, which is given by
\begin{equation}
T^{f} = \rho - 3p,
\label{35}
\end{equation}
into the on-shell classical matter action (\ref{24}), we obtain
\begin{equation}
S^{\textrm{on-shell}}_{m} = \frac{1}{4} \int{d^{4}x} \, \sqrt{-g} \, 
(3p - \rho),
\label{36}
\end{equation}
where $\rho$ and $p$ are the energy density and pressure of the perfect fluid 
that fills the universe. 

We can consider that all particles filling the universe are 
relativistic at early times. In this way, by substituting the pressure 
$p = \rho/3$ of a perfect fluid made up of such particles into 
(\ref{36}), we obtain
\begin{equation}
S^{\textrm{on-shell}}_{m} = 0.
\label{37}
\end{equation}
Since both (\ref{34}) and (\ref{37}) are finite, we conclude that the big bang 
singularity has no harmful effect on the cosmological quantum observables of 
MCG.


\section{Black hole singularity}
\label{sec5}


Let us consider that the spacetime geometry in the exterior of a static 
spherically symmetric mass distribution is described by the standard line 
element
\begin{equation}
ds^{2} = - B(r)dt^{2} + A(r)dr^2 + r^2\left(d\theta^2 
+ \sin^2\theta d\phi^2\right).
\label{38}
\end{equation}
The only known exact analytical static spherically symmetric solution to 
the MCG field equations (\ref{15}) and (\ref{16}) in vacuum ($T_{\mu\nu} = 0$) 
is the Schwarzschild black hole solution
\begin{equation}
B(r) = \frac{1}{A(r)} = 1 - \frac{r_0}{r},
\label{39}
\end{equation}
where $r_0$ is the radius of the black hole event horizon. However, 
this solution does not couple properly to a positive energy distributional 
source in MCG, as shown in Ref. \cite{Lu}\footnote{The 
vacuum field equations of MCG, and consequently its solutions, are 
the same as those of the Einstein-Weyl gravity studied in Ref. \cite{Lu}.}. 
Therefore, contrary to what happens in GR, the gravitational collapse of 
realistic physical matter will never form a Schwarzschild black 
hole in MCG and so analyzing its singularity here is meaningless.

Fortunately, there is a numerical horizonless black hole solution 
to (\ref{15}) and (\ref{16}) in vacuum which behave like a Schwarzschild 
black hole far away from the origin and fits precisely with a positive-energy 
distributional source \cite{Lu}. For simplicity, we will call this solution
a Lü-Perkins-Pope-Stelle (LPPS) black hole. Near the origin, which is our 
region of interest, the LPPS black hole is described by the Frobenius series
\begin{eqnarray}
A(r) &=& a_2 r^2 + a_2 b_3 r^3 - \frac{1}{6}a_2(b_3^2 + 2a_2 - 8b_4)r^4
- \frac{1}{54b_3}(-4a_2 b_3^4 + 33a_2^2 b_3^2 \nonumber \\ && 
- 75a_2 b_3^2 b_4 + 40b_3^4 + 30a_2^3 - 30a_2 b_4^2 + 135a_2^2 m_{g}^{2} )r^5 
+ \mathcal{O}(r^6),
\label{40}
\end{eqnarray}
\begin{eqnarray}
\frac{B(r)}{b_2} &=& r^2 + b_3 r^3 + b_4 r^4  
- \frac{1}{36a_2b_3}(2a_2^2 b_3^2 - 38a_2 b_3^2 b_4 + 16b_3^4 + 12a_2^3 
\nonumber \\ &&  - 12a_2b_4^2 + 54a_2^2 m_{g}^{2})r^5 + \mathcal{O}(r^6),
\label{41}
\end{eqnarray}
where $a_2$, $b_2$, $b_3$ and $b_4$ are non-vanishing coefficients 
and $m_{g} = \varphi_0\alpha$ is the mass of the spin-$2$ ghost.

The leading behaviors of the curvature invariants are given by
\begin{eqnarray}
R^{\alpha\beta\mu\nu}R_{\alpha\beta\mu\nu} &\sim& \frac{24}{a_2^2r^8} 
- \frac{32b_3}{a_2^2r^7}  + \mathcal{O}(r^{-6}), \label{42} \\
R^{\mu\nu}R_{\mu\nu} &\sim& \frac{12}{a_2^2r^8} - \frac{16b_3}{a_2^2r^7} 
+ \mathcal{O}(r^{-6}),
\label{43}
\end{eqnarray}
for (\ref{40}) and (\ref{41}), which means that the LPPS black hole is 
singular at the origin $r=0$. The fact that the LPPS black hole, which is 
the only other known singular solution with physical relevance of MCG, is
horizonless means that its naked singularity is visible to asymptotic observers 
so that it would be possible to see the physical signatures of the 
quantum gravity effects that may occur near such singularity. 

Just to finish our analysis, by substituting (\ref{42}) and (\ref{43}) 
into (\ref{17}), as well as the vacuum value $T^{f} = 0$ into (\ref{24}), 
we find that
\begin{equation}
S^{\textrm{on-shell}}_{\textrm{MCG}} = S^{\textrm{on-shell}}_{m} = 0.
\label{44}
\end{equation}
Therefore, just like the big bang singularity, the LPPS black 
hole singularity is also harmless in MCG.


\section{Conclusions}
\label{sec6}


We consider in this paper that for a renormalizable field theory to be 
consistent at the quantum level it is necessary that its on-shell classical 
action be finite at spacetime singularities. Based on this principle, we 
analyzed the behavior of the on-shell classical total action of MCG, which 
we have demonstrated to be a renormalizable theory of gravity, in singular 
spacetimes. It was shown that the MCG on-shell classical total action vanishes 
for the only two known singular solutions with physical relevance of the 
theory, namely the MCG universe and the LPPS black hole solutions. This, 
together with the renormalizability and unitarity of the theory, makes MCG 
a consistent quantum theory of gravity regardless of whether some of its 
classical spacetime solutions present singularities.

It would be important to check if other possible singular solutions with 
physical relevance of MCG lead to finite on-shell classical actions. However, 
finding such solutions is a very difficult task to do because the field 
equations of the theory are fourth-order nonlinear differential equations. 
Thus, we will leave such analyses for future works. Another topic that would 
be interesting to study in the future is how the gravitational field 
contributes to the entropy of the MCG universe. This is because the results 
obtained here contradicts the idea proposed by Penrose in Ref. \cite{Pen2} 
that the gravitational entropy is related to the Weyl curvature. We can see 
this contradiction by noting that the Weyl curvature invariant vanishes both 
at the initial big bang singularity and at the black hole singularities that 
arise later in the MCG universe due to gravitational collapse.


\end{document}